# Virtualization Architecture for NoC-based Reconfigurable Systems

Chun-Hsian Huang[1,†], Kwuan-Wei Tseng[2], Chih-Cheng Lin[2], Fang-Yu Lin[2], and Pao-Ann Hsiung[2]

[1]Department of Computer Science and Information Engineering, National Taitung University, Taiwan

[2]Deptartment of Computer Science and Information Engineering, National Chung Cheng University, Taiwan

[†]Email: huangch@nttu.edu.tw

## I. Introduction

To further enhance the capacity of parallel processing, the Network-on-Chip (NoC) is gradually adopted in a System-on-Chip (SoC) design, instead of the conventional bus architecture. Further, due to the support of partial reconfiguration technology, the Partial Reconfigurable Regions (PRRs) in an FPGA device can be configured as an IP core, such as a General-Purpose Processor (GPP) or a hardware accelerator. As a result, the Processing Elements (PEs) can be dynamically reconfigured on-demand in an NoC-based reconfigurable systems [1]. However, although the partial reconfiguration technology enhances system flexibility so as to meet different application requirements, the resource utilization of hardware logic is still restricted, owing to the limitation of NoC-based infrastructure. This means that, when a software application task is mapped to a PE, this PE is thus blocked and cannot be used by other application tasks, until the previous application finishes. In fact, the PE used is not accessed by the application all the time, which leads to a waste of computing resources.

To solve the above issue, we propose a virtualization architecture for NoC-based reconfigurable systems. The motivation of this work is to develop a service-oriented architecture that includes Partial Reconfigurable Region as a Service (PRRaaS) and Processing Element as a Service (PEaaS) for software applications. According to the requirements of software applications, new PEs can be created on-demand by (re)configuring the logic resource of the PRRs in the FPGA, while the configured PEs can also be virtualized to support multiple application tasks at the same time. As a result, such a two-level virtualization mechanism, including the gate-level virtualization and the PE-level virtualization, enables an SoC to be dynamically adapted to changing application requirements. Therefore, more software applications can be performed, and system performance can be further enhanced.

## II. Virtualization Architecture Design

The proposed design is based on a 2D-mesh architecture [2], as shown in Fig. 1. Different from the virtual channel design [3] that focuses on reducing congestion on an NoC, this work further introduces the concept of virtualization. In our current implementation, each PE can be virtualized as two virtual PEs to support two application tasks at the same time.

Besides adopting the partial reconfiguration flow to realize the PRRaaS, a new NI design and a new router design are proposed to realize the PEaaS, as shown in Fig. 2. To enable two different application tasks to access the same PE, in a router, two local ports are implemented to individually connect to the two DataReceive components (`DR0` and `DR1`) in the NI for supporting the PE-level virtualization. The DataReceive component is responsible for receiving the flits from a router, and all the received flits are then reconstructed to be a complete packet. When the reconstruction process of the packet finishes, the DataReceive component thus invokes the corresponding signal `AvReceive` (asserted high). Finally, the packet is transferred to the PE through the buffer controller using the signal `Data_in_0` or `Data_in_1`.

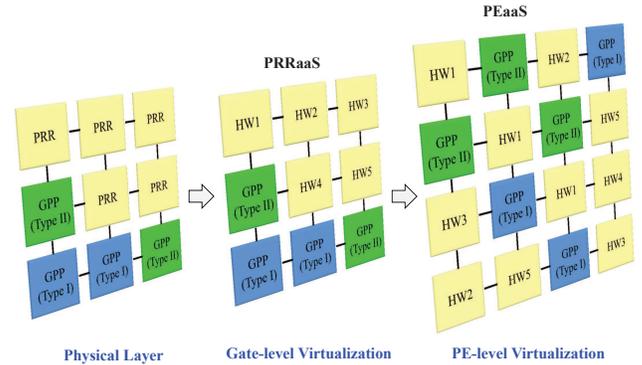

Fig. 1. Virtualization Architecture

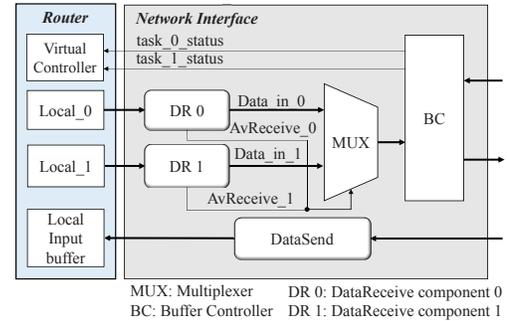

Fig. 2. Network Interface Design

To support the PE-level virtualization, each router also includes a virtualization controller that contains two specific signals, namely `task_0_status` and `task_1_status` to control the virtualization mechanism. Initially, the router would act as a conventional one that performs only an application task, in which one of the two local port is disabled. When a

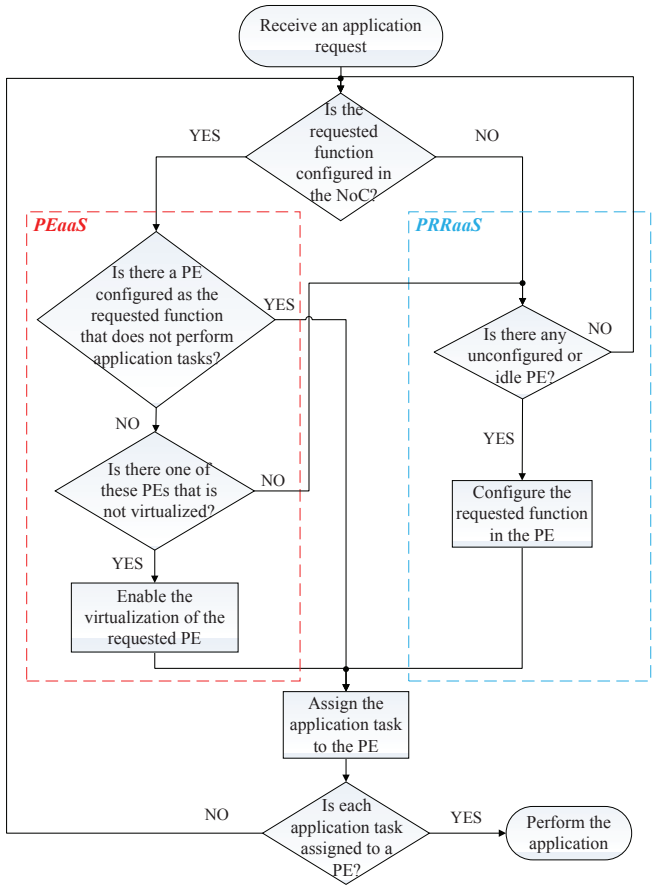

Fig. 3. Adaptation Management

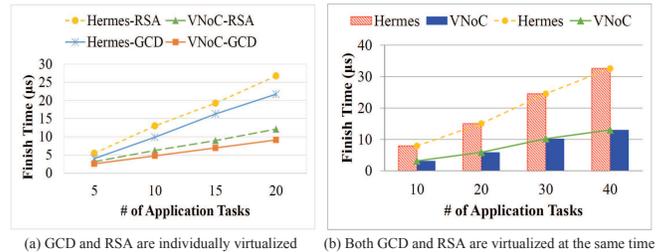

(a) GCD and RSA are individually virtualized

(b) Both GCD and RSA are virtualized at the same time

Fig. 4. System Performance Analysis

new application task needs to execute on the same PE, PE-level virtualization is then invoked. The unused local port is enabled to receive the packet flits from another application task. As a result, the flits received from two different application tasks can be individually and simultaneously transferred to the two DataReceive components. When the signal `AvReceive_0` or the signal `AvReceive_1` is asserted high, the corresponding signal (`Data_in_0` or `Data_in_1`) is thus used to transfer a complete packet to the PE. Here, based on the first-come-first-served scheduling policy, two different application tasks can be executed on the PE in an interleaving way. As a result, from the viewpoints of software applications, the PE is virtualized as two PEs.

To support the virtualization architecture, an adaptation management mechanism, as illustrated in Fig. 3 is also proposed to receive application requests. This mechanism is realized as a software program executed on a specific PE called global manager. By interfacing with the virtualization architecture, the PRRaaS and PEaaS can be performed for software applications.

## III. EXPERIMENTS

We implemented the virtualization design as a $3 \times 3$ mesh NoC architecture on the Xilinx Virtex 6 FPGA. To evaluate the proposed design, a conventional Hermes NoC design [2] was also implemented for comparison. Two PEs, including a RSA function and a Greatest Common Divisor (GCD) function, were used to execute multiple application tasks. Compared to the Hermes NoC, supporting PE virtualization needs an extra 1% of slice registers and an extra 2% of slice LUTs. The resource overheads in terms of additional reconfigurable resources are small and acceptable.

To evaluate performance improvement, different numbers of application tasks were applied to both the Hermes NoC and the proposed design (VNoC). The GCD function, the RSA function, and both the GCD and RSA functions were virtualized to support different numbers of application tasks, as shown in Figures 4(a) and 4(b), respectively. We can observe that, when the number of application tasks increases, performance improvement becomes more and more significant. This is because, through the support of the virtualization mechanism, a PE is no longer blocked by only an application task, and it can be used interleavingly by the two application tasks. However, in a conventional NoC, when an application task is mapped to a PE, another application task cannot be mapped to this PE, even though the PE is not used by the application all the time. According to our experimental results, the VNoC can accelerate by 1.5x to 2.5x the processing time required by using the conventional NoC design.

## IV. CONCLUSION AND FUTURE WORK

This work proposes an NoC-based virtualization design, which also provides the support of PRRaaS and PEaaS for software applications. By using this proposed design, both the utilization of system resources and system performance can be further enhanced. All the hardware adaptation processes are abstracted for the software applications and managed by the global manager, and thus software programmers can focus on the development of applications. In the next phase, we will extend the PE-level virtualization to support at most four software tasks. Further, the machine-learning method will be integrated into the adaptation management mechanism to provide a more intelligent management.